\documentclass{aa}

\usepackage[T1]{fontenc}
\usepackage{pslatex}
\usepackage{amsmath}
\usepackage{amsfonts}
\usepackage{xcolor}
\usepackage{url}
\usepackage{bm}
\usepackage{graphicx}
\usepackage{amssymb}
\usepackage{soul}
\usepackage{booktabs}
\usepackage{array}
\usepackage[utf8]{inputenc}
\inputencoding{latin1}
\inputencoding{utf8}
\usepackage{appendix}

\def\lsim{ \lower .75ex\hbox{$\sim$} \llap{\raise .27ex \hbox{$<$}} }
\def\gsim{ \lower .75ex \hbox{$\sim$} \llap{\raise .27ex \hbox{$>$}} }

\newcommand{\bi}{\begin{itemize}}
\newcommand{\ei}{\end{itemize}}

\newcommand{\pa}{\partial}

\usepackage[colorlinks=true,linkcolor=blue,citecolor=blue]{hyperref}

\title{Escape of fast radio bursts from magnetars}

\authorrunning{Sobacchi et al.}

\author{
E. Sobacchi\inst{1,2}
\and
M. Iwamoto\inst{3,4}
\and
L. Sironi\inst{5,6}
\and
T. Piran\inst{7}
}

\institute{
Gran Sasso Science Institute, viale F.~Crispi 7, L’Aquila 67100, Italy\\
\email{emanuele.sobacchi@gssi.it}
\and
INFN -- Laboratori Nazionali del Gran Sasso, via G.~Acitelli 22, Assergi 67100, Italy
\and
Yukawa Institute for Theoretical Physics, Kyoto University, Kitashirakawa-Oiwakecho, Sakyo-Ku, Kyoto 606-8502, Japan
\and
Faculty of Engineering Sciences, Kyushu University, 6-1, Kasuga-koen, Kasuga, Fukuoka 816-8580, Japan
\and
Department of Astronomy and Columbia Astrophysics Laboratory, Columbia University, 550 W 120th St, New York, NY 10027, USA
\and
Center for Computational Astrophysics, Flatiron Institute, 162 5th Avenue, New York, NY 10010, USA
\and
Racah Institute for Physics, The Hebrew University, Jerusalem, 91904, Israel
}
\date{}

\begin{document}


\abstract{
Fast radio bursts (FRBs) are bright extragalactic transients likely produced by magnetars. We study the propagation of FRBs in magnetar winds, assuming that the wind is strongly magnetized and composed of electron-positron pairs. We focus on the regime where the strength parameter of the radio wave, $a_0$, is larger than unity, and the wave frequency, $\omega_0$, is larger than the Larmor frequency in the background magnetic field, $\omega_{\rm L}$. We show that strong radio waves with $a_0>1$ are able to propagate when $\omega_0 > a_0\omega_{\rm L}$, as the plasma current is a linear function of the wave electric field. The dispersion relation is independent of the wave strength parameter when $\omega_0 > a_0\omega_{\rm L}$. Instead, radio waves could be damped when $\omega_0 < a_0\omega_{\rm L}$, as a significant fraction of the wave energy is used to compress the plasma and amplify the background magnetic field. Our results suggest that FRBs should be produced at large distances from the magnetar (i.e.,~$R>10^{12}{\rm\; cm}$, where the condition $\omega_0 > a_0\omega_{\rm L}$ is satisfied).  Alternatively, the structure of the magnetar wind should be strongly modified during a flare to allow the escape of FRBs produced  at radii $R<10^{12}{\rm\; cm}$.
}


\maketitle
\boldsymbol{}

\section{Introduction}

Fast radio bursts (FRBs) are bright extragalactic transients of millisecond duration \citep{CordesChatterjee2019, Petroff+2019, Petroff+2022}. Observationally, it is now clear that FRBs can be produced by magnetars \citep{Bochenek+2020, Chime2020}. However, despite significant theoretical effort, the physical processes powering FRBs have not been conclusively identified \citep{Lyubarsky2021, Zhang2023}.

Any emission model must satisfy the basic constraint that FRBs can escape from the magnetar. Due to the extraordinary properties of FRBs, this is not a trivial requirement. One can define the FRB strength parameter as
\begin{equation}
a_0 = \frac{eE_0}{\omega_0mc}\;,
\end{equation}
where $E_0$ is the peak electric field of the radio wave, $\omega_0$ is the wave frequency, $m$ and $-e$ are respectively the mass and the charge of the electron, and $c$ is the speed of light. When the background magnetic field is smaller than the field of the radio wave, the strength parameter is equal to the peak transverse component of the electrons four-velocity in units of the speed of light. Due to the large radio luminosity of FRBs, the strength parameter is $a_0>1$ at distances $R<10^{13}{\rm\; cm}$ from the magnetar \citep{LuanGoldreich2014}. It is an open question whether strong radio waves with $a_0>1$ can escape from the magnetar, or whether FRBs should be produced at large radii where $a_0<1$.

The magnetosphere and the wind of magnetars are strongly magnetized, and are composed primarily of electron-positron pairs \citep{KaspiBeloborodov2017}. The propagation of strong radio waves with $a_0>1$ in magnetized pair plasmas has been recently studied by several authors \citep{Beloborodov2021, Beloborodov2022, Beloborodov2023, Chen+2022, Qu+2022, GolbraikhLyubarsky2023, Lyutikov2024}. These studies focused on the regime where the wave frequency, $\omega_0$, is smaller than the Larmor frequency in the background magnetic field, $\omega_{\rm L}$. Two interesting cases have been identified: (i) When $\omega_0 < \omega_{\rm L}/a_0$, the particles oscillate in the field of the wave at nonrelativistic speeds because the background magnetic field is larger than the field of the radio wave. In this case, the plasma current is a linear function of the wave electric field, and radio waves can propagate. (ii) When $\omega_{\rm L}/a_0 < \omega_0 < \omega_{\rm L}$, the particles velocity is relativistic, and the plasma current is nonlinear. It has been suggested that in this case radio waves are damped, which would constrain emission models of FRBs \citep{Beloborodov2021, Beloborodov2022, Beloborodov2023, Chen+2022}. However, the latter issue is still debated \citep{Qu+2022, Lyutikov2024}.

In this paper, we study the propagation of strong waves in the regime $\omega_0>\omega_{\rm L}$, which has received limited attention so far.\footnote{\citet{Lyutikov2024} defines an ``effective nonlinearity parameter'', $\tilde{a}_0=a_0/(1+\omega_{\rm L}/\omega_0)$, which is of the order of the transverse component of the particles four-velocity. When $\omega_0<\omega_{\rm L}$ one has $\tilde{a}_0\sim a_0\omega_0/\omega_{\rm L}$, and when $\omega_0>\omega_{\rm L}$ one has $\tilde{a}_0\sim a_0$. The parameter $\tilde{a}_0$ is not the only controlling parameter of the problem because, as we show below, the plasma current can be a linear function of the wave electric field also when the particles velocity is relativistic.} In this regime, the particles velocity is relativistic. We show that: (i) When $\omega_0 > a_0\omega_{\rm L}$, the background magnetic field does not affect the particle motion, and the plasma current is a linear function of the wave electric field. Then, radio waves can propagate. Interestingly, the dispersion relation is independent of the wave strength parameter. (ii) When $\omega_{\rm L}<\omega_0<a_0\omega_{\rm L}$, the plasma current becomes nonlinear, and radio waves could be damped. The propagation regimes are summarized in Table \ref{tab:summary}.

The paper is organized as follows. In Sect.~\ref{sec:equations}, we present the fundamental equations that govern the propagation of strong waves. In Sect.~\ref{sec:super}, we consider waves that propagate with superluminal phase velocities, as appropriate for $\omega_0>\omega_{\rm L}$  (in Appendix \ref{sec:current}, we give an intuitive explanation of the results of Sect.~\ref{sec:equations}-\ref{sec:super}). In Sect.~\ref{sec:disc}, we discuss the implications of our results for the modelling of FRBs, and summarize our conclusions.

\section{Fundamental equations}
\label{sec:equations}

We consider a linearly polarized electromagnetic wave that propagates in a cold magnetized pair plasma. In the frame where the plasma ahead of the wave packet is at rest (hereafter, the ``lab frame'', or equivalently the proper frame of the magnetar wind), the wave frequency is $\omega_0$ and the wave vector is $k_0$. The wave propagates along the $z$ direction. We consider the extraordinary mode, where the electric field of the wave is directed along $y$, and the magnetic field of the wave is directed along $x$. The background magnetic field is also directed along $x$. In Sect.~\ref{sec:disc}, we argue that our key results are valid for any direction of propagation with respect to the background magnetic field, and for any direction of the wave electric field. We work in units where the speed of light is $c=1$. We assume that all physical quantities depend only on $z$ and $t$.

Electrons and positrons move in the $yz$ plane. They have the same number densities, the same velocities in the $z$ direction, and opposite velocities in the $y$ direction. The equation of motion of the positrons is
\begin{align}
\label{eq:eulery}
\frac{\pa u_y}{\pa t} + \frac{u_z}{\gamma}\frac{\pa u_y}{\pa z} & = \frac{e}{m}\left(E+\frac{u_z}{\gamma}B\right) \\
\label{eq:eulerz}
\frac{\pa u_z}{\pa t} + \frac{u_z}{\gamma}\frac{\pa u_z}{\pa z} & = -\frac{e}{m}\frac{u_y}{\gamma}B \;,
\end{align}
where ${\bm u} = u_y {\bm e}_y + u_z {\bm e}_z$ and $\gamma=\sqrt{1+{\bm u}^2}$ are respectively the four-velocity and the Lorentz factor. The electromagnetic fields, $E$ and $B$, are the sum of the wave field and the background field. The continuity equation is
\begin{equation}
\label{eq:cont}
\frac{\pa}{\pa t}\left(\gamma n\right) + \frac{\pa}{\pa z}\left(nu_z\right) = 0 \;,
\end{equation}
where $n$ is the proper number density. The evolution of the electromagnetic fields is governed by Faraday's and Amp\`ere's laws, which are
\begin{align}
\label{eq:faraday}
\frac{\pa E}{\pa z} & = \frac{\pa B}{\pa t} \\
\label{eq:ampere}
\frac{\pa B}{\pa z} & = 8\pi enu_y + \frac{\pa E}{\pa t} \;.
\end{align}
We define the plasma frequency as $\omega_{\rm P}=\sqrt{8\pi e^2n_0/m}$, where $n_0$ is the proper positron number density ahead of the wave packet, and the Larmor frequency as $\omega_{\rm L}=eB_{\rm bg}/m$, where $B_{\rm bg}$ is the background magnetic field in the lab frame. Since magnetar winds are strongly magnetized, we assume $\omega_{\rm L}>\omega_{\rm P}$.

It is convenient to apply the operator $\pa/\pa z$ to both sides of Eq.~\eqref{eq:eulery}. Using Eq.~\eqref{eq:faraday} to eliminate $\pa E/\pa z$, one finds
\begin{equation}
\label{eq:aux1}
\frac{\pa}{\pa t}\left(\frac{\pa u_y}{\pa z}-\frac{e}{m}B\right) + \frac{\pa}{\pa z}\left[\frac{u_z}{\gamma}\left(\frac{\pa u_y}{\pa z}-\frac{e}{m}B\right)\right] = 0 \;.
\end{equation}
Now one can introduce the new variable
\begin{equation}
\label{eq:Q}
Q= \frac{\pa u_y}{\pa z} - \frac{e}{m}B + \frac{\gamma n}{n_0}\omega_{\rm L}\;.
\end{equation}
Ahead of the wave packet one has $u_y=0$ and $n=n_0$. In the frame where the particles ahead of the wave packet are at rest, one has $\gamma=1$ and $(e/m)B=\omega_{\rm L}$. An arbitrary Lorentz boost along the direction of propagation gives $(e/m)B=\gamma\omega_{\rm L}$. Then, one finds $Q=0$ in any reference frame. From Eqs.~\eqref{eq:cont} and \eqref{eq:aux1}, one sees that the variable $Q$ is governed by
\begin{equation}
\label{eq:Qevo}
\frac{\pa Q}{\pa t} + \frac{\pa}{\pa z}\left(\frac{u_z}{\gamma} Q\right)=0\;.
\end{equation}
Then, one finds $Q=0$ also inside the wave packet. Substituting $Q=0$ into Eq.~\eqref{eq:Q}, one finds
\begin{equation}
\label{eq:B}
\frac{e}{m} B = \frac{\pa u_y}{\pa z} + \frac{\gamma n}{n_0}\omega_{\rm L}\;.
\end{equation}
Substituting Eq.~\eqref{eq:B} into Eq.~\eqref{eq:eulery}, one finds
\begin{equation}
\label{eq:E}
\frac{e}{m} E = \frac{\pa u_y}{\pa t} - \frac{n u_z}{n_0}\omega_{\rm L}\;.
\end{equation}
Substituting Eq.~\eqref{eq:B} into Eq.~\eqref{eq:eulerz}, one finds
\begin{equation}
\frac{\pa u_z}{\pa t} + \frac{u_y}{\gamma}\frac{\pa u_y}{\pa z} + \frac{u_z}{\gamma}\frac{\pa u_z}{\pa z} = -\frac{n u_y}{n_0}\omega_{\rm L} \;,
\end{equation}
or equivalently
\begin{equation}
\label{eq:final1}
\frac{\pa u_z}{\pa t} + \frac{\pa \gamma}{\pa z} = -\frac{n u_y}{n_0}\omega_{\rm L} \;.
\end{equation}
Substituting Eq.~\eqref{eq:B}-\eqref{eq:E} into Eq.~\eqref{eq:ampere}, one finds
\begin{equation}
\label{eq:final2}
\frac{\pa^2 u_y}{\pa t^2} - \frac{\pa^2 u_y}{\pa z^2} +\omega_{\rm P}^2\frac{n}{n_0}u_y = \omega_{\rm L}\left[\frac{\pa}{\pa t}\left(\frac{nu_z}{n_0}\right) + \frac{\pa}{\pa z}\left(\frac{\gamma n}{n_0}\right)\right] \;.
\end{equation}
Our final set of equations consists of Eqs.~\eqref{eq:cont}, \eqref{eq:final1}, \eqref{eq:final2}.

\begin{table}
\centering
\vspace{0.2cm}
\begin{tabular}{|c|c|c|}
 \hline
wave frequency & particles velocity & plasma current \\ [0.5ex] 
 \hline
$\omega_0<\omega_{\rm L}/a_0$ & nonrelativistic & linear \\ 
$\omega_{\rm L}/a_0<\omega_0<a_0\omega_{\rm L}$ & relativistic & nonlinear \\
$\omega_0>a_0\omega_{\rm L}$ & relativistic & linear \\
 \hline
\end{tabular}
\vspace{0.2cm}
\caption{Summary of the propagation regimes. The wave strength parameter is $a_0>1$. The plasma is strongly magnetized ($\omega_{\rm L}>\omega_{\rm P}$), and composed of electron-positron pairs. In the lab frame, the velocity of the particles that oscillate in the field of the wave is nonrelativistic for $\omega_0<\omega_{\rm L}/a_0$, and becomes relativistic for $\omega_0>\omega_{\rm L}/a_0$. The plasma current, $nu_y$, is a linear function of the wave electric field for $\omega_0<\omega_{\rm L}/a_0$ and $\omega_0>a_0\omega_{\rm L}$, whereas it is nonlinear for $\omega_{\rm L}/a_0<\omega_0<a_0\omega_{\rm L}$. The wave could be damped when the current is nonlinear.}
\label{tab:summary}
\vspace{-0.5cm}
\end{table}

\section{Superluminal waves}
\label{sec:super}

We study waves that propagate with superluminal phase velocities (i.e.,~$\omega_0/k_0>1$). We work in the reference frame that moves with the group velocity, $k_0/\omega_0$, along the $z$ direction (hereafter, the ``wave frame'').
As discussed by \citet{Clemmow1974}, in this frame the wave vector vanishes, and the wave frequency is
\begin{equation}
\omega=\sqrt{\omega_0^2-k_0^2} \;.
\end{equation}
We find an approximate solution of Eqs.~\eqref{eq:cont}, \eqref{eq:final1}, \eqref{eq:final2} by separating the fast oscillations of the physical quantities, which occur on the temporal scale $1/\omega$, from their secular evolution, which occurs on a much longer temporal scale.

Particles ahead of the wave packet move along the negative $z$ direction with a large four-velocity, $u_0=k_0/\omega\gg 1$. The four-velocity can be presented as
\begin{align}
\label{eq:uy}
u_y & = \delta u_{0y} +\frac{1}{2}u_{{\rm f}y}{\rm e}^{-{\rm i}\omega t}+\ldots+{\rm c.c.} \\
\label{eq:uz}
u_z & = -u_0 +\delta u_{0z} +\frac{1}{2}u_{{\rm f}z}{\rm e}^{-{\rm i}\omega t}+\ldots+{\rm c.c.}
\end{align}
The proper density can be presented as
\begin{equation}
\label{eq:n}
n = n_0 + \delta n_0 + \frac{1}{2}n_{\rm f} {\rm e}^{-{\rm i}\omega t}+\ldots+{\rm c.c.}
\end{equation}
We introduced the real functions $\delta {\bm u}_0 = \delta u_{0y}{\bm e}_y + \delta u_{0z}{\bm e}_z$ and $\delta n_0$, and the complex functions ${\bm u}_{\rm f} = u_{{\rm f}y}{\bm e}_y + u_{{\rm f}z}{\bm e}_z$ and $n_{\rm f}$. For example, $\delta {\bm u}_0(z,t)$ describes the secular evolution of the four-velocity inside the wave packet, and ${\bm u}_{\rm f}(z,t)$ describes the amplitude of the fast oscillations of the four-velocity. When the envelope of the wave packet is much longer than a few wavelengths, these functions vary on spatial and temporal scales $\gg 1/\omega$. The dots in Eqs.~\eqref{eq:uy}-\eqref{eq:n} indicate higher-order harmonics, and c.c.~indicates the complex conjugate of the oscillating terms. The expansion in harmonics is justified when the plasma current is a linear function of the wave electric field (i.e.,~for $\omega_0\gg a_0\omega_{\rm L}$, as we show below). Second-order harmonics would be important to calculate nonlinear corrections to the dispersion relation of the order of $u_{\rm f}^2/u_0^2$, which is out of the scope of the paper.

We consider the regime where the four-velocity inside the wave packet is nearly constant, and assume $\delta u_0/u_0\sim u_{\rm f}^2/u_0^2\ll 1$. Below we show that the condition $\delta u_0/u_0\sim u_{\rm f}^2/u_0^2\ll 1$ requires $\omega_0\gg a_0\omega_{\rm L}$. In this regime, the plasma current is a linear function of the wave electric field.

Retaining terms of the order of $\delta u_0/u_0$ and $u_{\rm f}^2/u_0^2$, the Lorentz factor can be presented as
\begin{equation}
\label{eq:gamma}
\frac{\gamma}{\gamma_0} = 1 - \frac{u_0}{\gamma_0}\frac{\delta u_{0z}}{\gamma_0} + \frac{\left|u_{{\rm f}y}\right|^2}{4\gamma_0^2} - \frac{u_0}{\gamma_0}\frac{u_{{\rm f}z}}{2\gamma_0} {\rm e}^{-{\rm i}\omega t} + \ldots + {\rm c.c.} \;,
\end{equation}
where $\gamma_0=\sqrt{1+u_0^2}=\omega_0/\omega$. The number density is
\begin{align}
\nonumber
\frac{\gamma n}{\gamma_0 n_0} = 1 + \frac{\delta n_0}{n_0} & - \frac{u_0}{\gamma_0}\frac{\delta u_{0z}}{\gamma_0} + \frac{\left|u_{{\rm f}y}\right|^2}{4\gamma_0^2} - \frac{u_0}{\gamma_0}\frac{n_{\rm f}u_{{\rm f}z}^*+n_{\rm f}^*u_{{\rm f}z}}{4\gamma_0n_0} + \\
\label{eq:N}
& + \frac{1}{2}\left(\frac{n_{\rm f}}{n_0}-\frac{u_0}{\gamma_0}\frac{u_{{\rm f}z}}{\gamma_0}\right){\rm e}^{-{\rm i}\omega t} + \ldots + {\rm c.c.}
\end{align}
The $y$ component of the current density is
\begin{equation}
\label{eq:jy}
\frac{nu_y}{\gamma_0 n_0} = \frac{\delta u_{0y}}{\gamma_0} + \frac{n_{\rm f}u_{{\rm f}y}^*+n_{\rm f}^*u_{{\rm f}y}}{4\gamma_0 n_0} + \frac{u_{{\rm f}y}}{2\gamma_0}{\rm e}^{-{\rm i}\omega t} + \ldots + {\rm c.c.}\;,
\end{equation}
and the $z$ component is
\begin{align}
\nonumber
\frac{nu_z}{\gamma_0 n_0} = -\frac{u_0}{\gamma_0} & -\frac{u_0}{\gamma_0}\frac{\delta n_0}{n_0} + \frac{\delta u_{0z}}{\gamma_0} + \frac{n_{\rm f}u_{{\rm f}z}^*+n_{\rm f}^*u_{{\rm f}z}}{4\gamma_0 n_0} + \\
\label{eq:jz}
& + \frac{1}{2}\left( \frac{u_{{\rm f}z}}{\gamma_0} -\frac{u_0}{\gamma_0}\frac{n_{\rm f}}{n_0} \right) {\rm e}^{-{\rm i}\omega t} + \ldots + {\rm c.c.} \;,
\end{align}
where $u_{\rm f}^*$ and $n_{\rm f}^*$ indicate respectively the complex conjugates of $u_{\rm f}$ and $n_{\rm f}$.

\subsection{Fast oscillations}
\label{sec:fastsuper}

When the envelope of the wave packet is much longer than a few wavelengths, the functions $\delta u_0$, $\delta n_0$, $u_{\rm f}$, $n_{\rm f}$ vary on spatial and temporal scales $\gg 1/\omega$. The evolution of the physical quantities on the temporal scale $1/\omega$ can be easily determined. Since in the wave frame the wave vector vanishes, one can neglect the spatial derivatives of the physical quantities in Eqs.~\eqref{eq:cont}, \eqref{eq:final1}, \eqref{eq:final2}. One can also approximate the time derivatives as $\pa/\pa t\simeq -{\rm i}\omega$. From Eq.~\eqref{eq:cont} one sees that $\gamma n$ should be constant on the temporal scale $1/\omega$. Then, Eq.~\eqref{eq:N} gives
\begin{equation}
\label{eq:nfast}
\frac{n_{\rm f}}{n_0} = \frac{u_0}{\gamma_0}\frac{u_{{\rm f}z}}{\gamma_0} \;.
\end{equation}
Substituting Eqs.~\eqref{eq:uz} and \eqref{eq:jy} into Eq.~\eqref{eq:final1}, one finds ${\rm i}\omega u_{{\rm f}z} = \omega_{\rm L}u_{{\rm f}y}$, or equivalently
\begin{equation}
\label{eq:uzfast}
u_{{\rm f}z} = - {\rm i} \frac{\omega_{\rm L}}{\omega} u_{{\rm f}y} \;.
\end{equation}
Substituting Eqs.~\eqref{eq:uy}, \eqref{eq:jy}, \eqref{eq:jz} into Eq.~\eqref{eq:final2} one finds
\begin{equation}
\label{eq:wavefast}
-\omega^2 u_{{\rm f}y} + \omega_{\rm P}^2 u_{{\rm f}y} = -{\rm i}\omega\omega_{\rm L} \left( u_{{\rm f}z} - u_0\frac{n_{\rm f}}{n_0} \right) \;.
\end{equation}
Substituting Eqs.~\eqref{eq:nfast}-\eqref{eq:uzfast} into Eq.~\eqref{eq:wavefast}, one finds
\begin{equation}
\label{eq:DRsuper}
\omega^2 \left( 1 - \frac{\omega_{\rm L}^2}{\gamma_0^2\omega^2} \right) = \omega_{\rm P}^2 \;.
\end{equation}
Taking into account that $\omega^2=\omega_0^2-k_0^2$ and $\gamma_0^2\omega^2=\omega_0^2$, one sees that Eq.~\eqref{eq:DRsuper} is equivalent to the standard dispersion relation in the lab frame, which is $(\omega_0^2-k_0^2)(1-\omega_{\rm L}^2/\omega_0^2)=\omega_{\rm P}^2$. The wave is superluminal for $\omega_0 \gg \omega_{\rm L}$.

Now we discuss the range of validity of our results. We derived the dispersion relation assuming that the particle four-velocity inside the wave packet is nearly constant (i.e.,~$|u_{{\rm f}y}|\ll \gamma_0$ and $|u_{{\rm f}z}|\ll \gamma_0$). For superluminal waves with $\omega_0 \gg \omega_{\rm L}$, one has $\gamma_0= \omega_0/\omega_{\rm P}$, $|u_{{\rm f}y}|= a_0$ and $|u_{{\rm f}z}|= (\omega_{\rm L}/\omega_{\rm P})|u_{{\rm f}y}|$ (the latter two relations follow from Eqs.~\ref{eq:E} and \ref{eq:uzfast}). Since the plasma is strongly magnetized, one has $|u_{{\rm f}z}| \gg |u_{{\rm f}y}|$. The condition $|u_{{\rm f}z}| \ll \gamma_0$ requires\footnote{In weakly magnetized pair plasmas where $\omega_{\rm L}\ll\omega_{\rm P}$, one has $|u_{{\rm f}z}| \ll |u_{{\rm f}y}|$. The condition $|u_{{\rm f}y}| \ll \gamma_0$ requires $\omega_0 \gg a_0\omega_{\rm P}$. As we explain in Appendix \ref{sec:current}, the latter condition implies that in the lab frame the particles velocity along the direction of propagation should be smaller than the group velocity of the wave.}
\begin{equation}
\label{eq:condsuper}
\omega_0 \gg a_0\omega_{\rm L} \;.
\end{equation}
Eq.~\eqref{eq:condsuper} has a simple physical interpretation. As we show in Sect.~\ref{sec:secsuper}, the plasma inside the wave packet is compressed, and consequently the background magnetic field is amplified. In the lab frame, the amplified magnetic field is of the order of $a_0^2B_{\rm bg}$. Then, Eq.~\eqref{eq:condsuper} implies that the amplified magnetic field should be smaller than the wave field. We emphasize that Eq.~\eqref{eq:condsuper} can be satisfied by strong waves with $a_0> 1$ when the wave frequency is much larger than the Larmor frequency.

\subsection{Secular evolution}
\label{sec:secsuper}

In order to study the secular evolution of the physical quantities, one should average Eqs.~\eqref{eq:cont}, \eqref{eq:final1}, \eqref{eq:final2} over the fast oscillations.  In this case, the spatial derivatives of the physical quantities should be retained. From Eq.~\eqref{eq:cont}, one finds
\begin{equation}
\label{eq:contave}
\frac{\pa}{\pa t}\left(\frac{\langle\gamma n\rangle}{\gamma_0n_0}\right) + \frac{\pa}{\pa z}\left(\frac{\langle nu_z\rangle}{\gamma_0n_0}\right) = 0 \;.
\end{equation}
Eq.~\eqref{eq:final1} gives
\begin{equation}
\label{eq:final1slow}
\frac{\pa}{\pa t}\left(\frac{\langle u_z\rangle}{\gamma_0}\right) + \frac{\pa}{\pa z}\left(\frac{\langle \gamma \rangle}{\gamma_0}\right) = -\frac{\langle nu_y \rangle}{\gamma_0n_0}\omega_{\rm L} \;.
\end{equation}
Eq.~\eqref{eq:final2} gives
\begin{align}
\nonumber
\frac{\pa^2}{\pa t^2}\left(\frac{\langle u_y \rangle}{\gamma_0}\right) & - \frac{\pa^2}{\pa z^2}\left(\frac{\langle u_y\rangle}{\gamma_0}\right) +\omega_{\rm P}^2\frac{\langle u_y \rangle}{\gamma_0} = \\
\label{eq:final2slow}
& = \omega_{\rm L} \left[ \frac{\pa}{\pa t}\left( \frac{\langle nu_z\rangle}{\gamma_0n_0}\right) + \frac{\pa}{\pa z}\left( \frac{\langle \gamma n\rangle}{\gamma_0n_0}\right) \right] \;.
\end{align}
In Appendix \ref{sec:average}, we present the expressions for the average physical quantities that appear in Eqs.~\eqref{eq:contave}-\eqref{eq:final2slow}.

We find an approximate solution of Eqs.~\eqref{eq:contave}-\eqref{eq:final2slow} by neglecting the time derivatives of the average physical quantities. This ``quasi-static'' approximation has been extensively used to study laser-plasma interaction \citep[e.g.,][]{Sprangle+1990}. The approximation is justified because we work in the frame that moves with the group velocity, where the time derivative of the wave envelope can be neglected. The average physical quantities reach a steady state because (as we show below) they can be expressed as a function of $a_0^2$. For example, since $\pa a_0^2/\pa t\ll\pa a_0^2/\pa z$ and $\langle u_y\rangle$ is a function of $a_0^2$, one has $\pa \langle u_y\rangle/\pa t\ll\pa \langle u_y\rangle/\pa z$.

The average quantities vary on a scale comparable to the length of the wave packet, which in the wave frame is much longer than $1/\omega_{\rm P}$. Then, for example, one has $\pa^2\langle u_y\rangle/\pa z^2\ll\omega_{\rm P}^2\langle u_y\rangle$. In this case, the third term on the left hand side of Eq.~\eqref{eq:final2slow} is much larger than the second term, and Eq.~\eqref{eq:final2slow} can be approximated as $\omega_{\rm P}^2\langle u_y\rangle/\gamma_0=\omega_{\rm L} \pa (\langle\gamma n\rangle/\gamma_0n_0)/\pa z$. Using Eqs.~\eqref{eq:uyave} and \eqref{eq:Nave}, the latter equation can be presented as
\begin{equation}
\label{eq:deltauysol}
\frac{\delta u_{0y}}{\gamma_0} = \frac{\omega_{\rm L}}{\omega_{\rm P}^2}\frac{\pa}{\pa z}\left(  \frac{\delta n_0}{n_0} - \frac{u_0}{\gamma_0}\frac{\delta u_{0z}}{\gamma_0} + \frac{a_0^2}{4\gamma_0^2} - \frac{u_0^2}{\gamma_0^2}\frac{\omega_{\rm L}^2}{\omega_{\rm P}^2}\frac{a_0^2}{2\gamma_0^2} \right) \;.
\end{equation}
Eq.~\eqref{eq:final1slow} can be approximated as $\pa (\langle\gamma\rangle/\gamma_0)/\pa z= -\omega_{\rm L} \langle nu_y \rangle/\gamma_0 n_0$. Substituting Eqs.~\eqref{eq:gammaave} and \eqref{eq:jyave} into the latter equation, and using Eq.~\eqref{eq:deltauysol} to eliminate $\delta u_{0y}$, one finds
\begin{align}
\nonumber
\frac{\pa}{\pa z} & \left(\frac{u_0}{\gamma_0}\frac{\delta u_{0z}}{\gamma_0} - \frac{a_0^2}{4\gamma_0^2}\right) = \\
& = \frac{\omega_{\rm L}^2}{\omega_{\rm P}^2}\frac{\pa}{\pa z}\left(  \frac{\delta n_0}{n_0} - \frac{u_0}{\gamma_0}\frac{\delta u_{0z}}{\gamma_0} + \frac{a_0^2}{4\gamma_0^2} - \frac{u_0^2}{\gamma_0^2}\frac{\omega_{\rm L}^2}{\omega_{\rm P}^2}\frac{a_0^2}{2\gamma_0^2} \right)\;,
\end{align}
which implies
\begin{equation}
\label{eq:sol1}
\frac{\delta n_0}{n_0} = \frac{u_0}{\gamma_0} \left(1+\frac{\omega_{\rm P}^2}{\omega_{\rm L}^2} \right)\frac{\delta u_{0z}}{\gamma_0} + \left(\frac{2u_0^2}{\gamma_0^2}\frac{\omega_{\rm L}^2}{\omega_{\rm P}^2} -1-\frac{\omega_{\rm P}^2}{\omega_{\rm L}^2}\right)\frac{a_0^2}{4\gamma_0^2} \;.
\end{equation}
Eq.~\eqref{eq:contave} can be approximated as $\pa \langle nu_z\rangle/\pa z=0$, which implies
\begin{equation}
\label{eq:jzavesol}
\langle nu_z\rangle=-n_0u_0\;.
\end{equation}
Substituting Eq.~\eqref{eq:jzavesol} into Eq.~\eqref{eq:jzave}, one finds
\begin{equation}
\label{eq:sol2}
\frac{\delta n_0}{n_0} = \frac{\gamma_0}{u_0}\frac{\delta u_{0z}}{\gamma_0} + \frac{\omega_{\rm L}^2}{\omega_{\rm P}^2}\frac{a_0^2}{2\gamma_0^2} \;.
\end{equation}

Now one can solve Eqs.~\eqref{eq:sol1} and \eqref{eq:sol2}, and express $\delta u_{0z}$ and $\delta n_0$ as a function of $a_0^2$. Then, one can substitute $\delta u_{0z}$ and $\delta n_0$ into Eqs.~\eqref{eq:uzave}, \eqref{eq:gammaave} and \eqref{eq:Nave}, and express also $\langle u_z\rangle$, $\langle\gamma\rangle$ and $\langle\gamma n\rangle$ as a function of $a_0^2$. This procedure gives
\begin{align}
\label{eq:uzavesol}
\langle u_z\rangle & =  - u_0 + \left(1+\frac{\omega_{\rm L}^2}{\omega_{\rm P}^2}\right)\frac{a_0^2}{4u_0}  \\
\label{eq:gammaavesol}
\langle \gamma\rangle & = \gamma_0 - \frac{\omega_{\rm L}^2}{\omega_{\rm P}^2} \frac{a_0^2}{4\gamma_0} \\
\label{eq:Navesol}
\langle \gamma n\rangle & = \left(\gamma_0 +\frac{a_0^2}{4\gamma_0} \right) n_0 \;.
\end{align}

The average four-velocity and the average Lorentz factor in the lab frame can be determined from Eqs.~\eqref{eq:uzavesol}-\eqref{eq:gammaavesol}. Taking into account that $\omega_0\gg a_0\omega_{\rm L}$, one finds
\begin{align}
\label{eq:uzlab}
\langle u_z \rangle_{\rm lab} & = \frac{a_0^2}{4} \\
\label{eq:gammalab}
\langle\gamma\rangle_{\rm lab} & = 1 + \frac{a_0^2}{4} \;.
\end{align}
The background magnetic field does not affect the particle motion in the lab frame, as Eqs.~\eqref{eq:uzlab}-\eqref{eq:gammalab} are identical to the case of a test particle that moves in the field of a vacuum wave \citep[e.g.,][]{GunnOstriker1971}. 

The average current density and the average number density in the lab frame can be determined from Eqs.~\eqref{eq:jzavesol} and \eqref{eq:Navesol}. One finds
\begin{align}
\label{eq:jzlab}
\langle nu_z \rangle_{\rm lab} & = \frac{a_0^2}{4} n_0 \\
\label{eq:Nlab}
\langle\gamma n \rangle_{\rm lab} & = \left(1 + \frac{a_0^2}{4}\right) n_0 \;.
\end{align}
Substituting Eqs.~\eqref{eq:jzlab}-\eqref{eq:Nlab} into Eqs.~\eqref{eq:B}-\eqref{eq:E}, one can determine the average electromagnetic fields in the lab frame, which are given by
\begin{align}
\label{eq:Elab}
\langle E \rangle_{\rm lab} & = \frac{a_0^2}{4} B_{\rm bg} \\
\label{eq:Blab}
\langle B \rangle_{\rm lab} & = \left( 1+ \frac{a_0^2}{4} \right) B_{\rm bg} \;.
\end{align}
In the lab frame, the number density and the background magnetic field inside the wave packet are amplified by a large factor of the order of $a_0^2$. The amplified magnetic field is smaller than the wave field when $\omega_0\gg a_0\omega_{\rm L}$. When the latter condition is violated, particles could be trapped inside the wave packet because $\langle u_z\rangle$ could vanish in the wave frame (see Eq.~\ref{eq:uzavesol}).

\section{Discussion}
\label{sec:disc}

We studied the propagation of superluminal electromagnetic waves with strength parameters $a_0> 1$ in magnetized pair plasmas where $\omega_{\rm L}>\omega_{\rm P}$. Our key result is that strong waves are able to propagate when $\omega_0>a_0\omega_{\rm L}$, where $\omega_0$ is the wave frequency, and $\omega_{\rm L}$ is the Larmor frequency in the background magnetic field. When $\omega_0 > a_0\omega_{\rm L}$, the background magnetic field does not affect the particle motion, and the plasma current is a linear function of the wave electric field. The dispersion relation is independent of the wave strength parameter.

Strong superluminal waves could be damped when $\omega_{\rm L}<\omega_0< a_0\omega_{\rm L}$, for the following reasons. (i) The plasma inside the wave packet is compressed, and its density is amplified by a large factor of the order of $a_0^2$. Although we demonstrated this for the extraordinary mode, we expect that the density is amplified also by the ordinary mode, as the particle motion is not affected by the background field, and is nearly identical to the case of a test particle in the field of a vacuum wave. (ii) Since the background magnetic field is frozen into the plasma, its strength is also amplified by a factor of the order of $a_0^2$. When $\omega_0< a_0\omega_{\rm L}$, the amplified background field would be larger than the wave field. Then, the wave would be damped, as a significant fraction of its energy would be used to compress the plasma and amplify the background field. The only exception may occur when the wave propagates nearly along the background field, as the background field would not be significantly amplified. This exception does not apply to magnetar winds, where the magnetic field is nearly toroidal, and the wave propagates in the radial direction. Fully-kinetic simulations can be used to study the damping of strong superluminal waves when $\omega_{\rm L}<\omega_0< a_0\omega_{\rm L}$.

Our results have important implications for the modelling of FRBs. Let us first summarize the typical parameters of the problem \citep[see, e.g.,][]{Sobacchi+2022}. The FRB strength parameter is
\begin{equation}
\label{eq:aFRB}
a_0 = 20 \; L_{42}^{1/2} \nu_9^{-1} R_{12}^{-1} \;,
\end{equation}
where $L=10^{42}L_{42}{\rm\; erg\; s^{-1}}$ is the isotropic equivalent FRB luminosity, $\nu = 1\; \nu_9 {\rm\; GHz}$ is the observed FRB frequency, and $R=10^{12}R_{12}{\rm\; cm}$ is the distance from the magnetar. The magnetar light cylinder is located at the radius $R_{\rm lc}=cP/2\pi$, where $P=1\; P_0{\rm\; s}$ is the magnetar period. In the magnetosphere (i.e.,~for $R<R_{\rm lc}$), the magnetar has a dipole field, $B_{\rm bg}=\mu/R^3$, where $\mu=10^{33}\mu_{33}{\rm\; G\; cm^3}$ is the magnetic moment. In the wind (i.e.,~for $R>R_{\rm lc}$), the magnetic field is $B_{\rm bg}=\mu/R_{\rm lc}^2R$. In the proper frame of the wind (which we called ``lab frame'' so far), the ratio of the Larmor frequency and the FRB frequency is
\begin{equation}
\label{eq:omegaratio}
\frac{\omega_{\rm L}}{\omega_0} = 0.2 \; \mu_{33} \nu_9^{-1} P_0^{-2} R_{12}^{-1} \;.
\end{equation}
Eqs.~\eqref{eq:aFRB}-\eqref{eq:omegaratio} are independent of the wind Lorentz factor because $a_0$ and $\omega_{\rm L}/\omega_0$ are Lorentz invariant.

Now we show that FRBs can propagate through the magnetar wind only at relatively large radii.\footnote{Subluminal waves with $a_0>1$ can propagate when $\omega_0<\omega_{\rm L}/a_0$, whereas they could be damped when $\omega_{\rm L}/a_0< \omega_0 <\omega_{\rm L}$ \citep{Beloborodov2021, Beloborodov2022, Beloborodov2023, Chen+2022}. The condition $\omega_0<\omega_{\rm L}/a_0$ can be satisfied only well within the magnetar magnetosphere, as it requires $R<4\times 10^8L_{42}^{-1/4}\mu_{33}^{1/2}{\rm\; cm}$ \citep{Beloborodov2021}.} The condition $\omega_0>a_0\omega_{\rm L}$ requires $R> R_{\rm crit}$, where
\begin{equation}
R_{\rm crit} = 2 \times 10^{12} L_{42}^{1/4} \mu_{33}^{1/2} \nu_9^{-1} P_0^{-1} {\rm\; cm} \;.
\end{equation}
Two consistency checks are needed. Since we considered the propagation of the wave in the wind, $R_{\rm crit}$ should be larger than the light cylinder radius. The condition $R_{\rm crit}>R_{\rm lc}$ is satisfied for
\begin{equation}
\label{eq:P1}
P < 20 \; L_{42}^{1/8} \mu_{33}^{1/4} \nu_9^{-1/2} {\rm\; s} \;.
\end{equation}
Since we assumed that the strength parameter is large, one should have $a_0>1$ for $R=R_{\rm crit}$. This condition is satisfied for
\begin{equation}
\label{eq:P2}
P > 0.1\; L_{42}^{-1/4} \mu_{33}^{1/2} {\rm\; s} \;.
\end{equation}
Eqs.~\eqref{eq:P1}-\eqref{eq:P2} are satisfied by most magnetars, which have a typical period of a few seconds \citep{KaspiBeloborodov2017}.

Our results suggest that FRBs produced at radii $R< R_{\rm crit}$ cannot propagate through the unperturbed magnetar wind, which challenges several magnetospheric models \citep[e.g.,][]{CordesWasserman2016, Dai+2016, Kumar+2017, YangZhang2018, LuKumar2018, KumarBosnjak2020, Lu+2020}. There are two alternative possibilities: (i)~FRBs could be produced at radii $R> R_{\rm crit}$, as in the shock maser model \citep{Lyubarsky2014, Beloborodov2017, Beloborodov2020, Metzger+2019, Sironi+2021, Iwamoto+2024}. (ii)~Alternatively, the structure of the wind could be modified during a flare, as in the reconnection model \citep{Lyubarsky2020, Mahlmann+2022}. In this model, FRBs produced at radii $R< R_{\rm crit}$ propagate on the top of a large amplitude fast-magnetosonic pulse launched from the magnetosphere. FRBs can escape because their electromagnetic field is much smaller than the field of the pulse.

\begin{acknowledgements}
This work was supported by the JSPS KAKENHI Grants 20J00280, 20KK0064, and 22H00130 [M.I.], by the Simons Foundation Grant 00001470 to the Simons Collaboration on Extreme Electrodynamics of Compact Sources (SCEECS) [L.S., T.P.], by the DoE Early Career Award DE-SC0023015 [L.S.], by the Multimessenger Plasma Physics Center (MPPC) NSF Grant PHY-2206609 [L.S.], by the NASA Grant 23-ATP23-0074 [L.S.], by the ISF Grant 2126/22 [T.P.], and by the ERC Advanced Grant Multijets [T.P.]. We acknowledge insightful discussions with Andrei Beloborodov and Yuri Lyubarsky.
\end{acknowledgements}

\bibliographystyle{aa}
\bibliography{2d}

\appendix
\section{Test particles in a vacuum wave}
\label{sec:current}

The motion of test particles in the electromagnetic fields of a strong vacuum wave has been studied for a long time. Here we summarize the main results of previous studies \citep[e.g.,][]{GunnOstriker1971}. We work in the lab frame where the particles ahead of the wave packet are at rest. We set $\omega_{\rm L}=0$ in Eqs.~\eqref{eq:cont}, \eqref{eq:final1}, \eqref{eq:final2}. We also neglect the feedback of plasma current onto the wave, and set $\omega_{\rm P}=0$ in Eq.~\eqref{eq:final2}. In this case, the $y$ component of the four-velocity can be presented as
\begin{equation}
\label{eq:uyapp}
u_y = a_0\cos\left(\omega_0\xi\right) \;,
\end{equation}
where $\xi=z-t$, and $a_0>1$ is the wave strength parameter.

Eq.~\eqref{eq:final1} can be presented as ${\rm d}(\gamma-u_z)/{\rm d}\xi=0$. In the lab frame where the particles ahead of the wave packet are at rest, the latter equation has the solution $\gamma-u_z=1$, which gives
\begin{equation}
\gamma =1+\frac{u_y^2}{2} \;, \qquad u_z = \frac{u_y^2}{2} \;.
\end{equation}
Taking into account that $\langle u_y^2\rangle = a_0^2/2$, one recovers Eqs.~\eqref{eq:uzlab}-\eqref{eq:gammalab}. Eq.~\eqref{eq:cont} can be presented as ${\rm d}[(\gamma-u_z)n]/{\rm d}\xi=0$, where $n$ is the proper number density. Taking into account that $\gamma-u_z=1$, one finds
\begin{equation}
\label{eq:napp}
n=n_0\;,
\end{equation}
where $n_0$ is the proper number density ahead of the wave packet. Eqs.~\eqref{eq:uyapp} and \eqref{eq:napp} show that in the test particle approximation the plasma current, $nu_y$, is a linear function of the wave field.

Now we discuss the validity of the test particle approximation. We focus on pair plasmas, where an electrostatic field is not generated. As we demonstrate formally in Sect.~\ref{sec:super}, two key conditions should be satisfied. First, the particles velocity along the direction of propagation in the lab frame, $v_z=1-2/(2+u_y^2)$, should be smaller than the group velocity of the wave, $v_{\rm g}\simeq 1-\omega_{\rm P}^2/2\omega_0^2$, otherwise particles would be trapped inside the wave packet. This condition is satisfied for
\begin{equation}
\omega_0\gg a_0\omega_{\rm P}\;.
\end{equation}
Second, the average magnetic field inside the wave packet in the lab frame, $\langle B\rangle$, should be smaller than the peak electric field of the wave, $E_0=(m/e)a_0\omega_0$. The particles number density inside the wave packet is equal to $\gamma n_0$, whereas the particles number density ahead of it is equal to $n_0$. Since the average magnetic field is frozen into the plasma, one has $\langle B\rangle = \langle\gamma\rangle B_{\rm bg} = (1+a_0^2/4)B_{\rm bg}$, where $B_{\rm bg}$ is the strength of the background magnetic field ahead of the wave packet. The average electric field is $\langle E\rangle=\langle v_z B \rangle=(a_0^2/4)B_{\rm bg}$ (where $v_z=u_z/\gamma$). The condition $\langle B\rangle \ll E_0$ is satisfied for
\begin{equation}
\label{eq:cond2app}
\omega_0\gg a_0\omega_{\rm L}\;.
\end{equation}
In strongly magnetized plasmas where $\omega_{\rm L}\gg\omega_{\rm P}$, the current is a linear function of the wave field when Eq.~\eqref{eq:cond2app} is satisfied.

\section{Average physical quantities}
\label{sec:average}

The average four-velocity and the average Lorentz factor can be calculated from Eqs.~\eqref{eq:uy}, \eqref{eq:uz}, \eqref{eq:gamma}, which give respectively
\begin{align}
\label{eq:uyave}
\frac{\langle u_y\rangle}{\gamma_0} & = \frac{\delta u_{0y}}{\gamma_0} \\
\label{eq:uzave}
\frac{\langle u_z\rangle}{\gamma_0} & = -\frac{u_0}{\gamma_0} + \frac{\delta u_{0z}}{\gamma_0} \\
\frac{\langle \gamma\rangle}{\gamma_0} & = 1 - \frac{u_0}{\gamma_0}\frac{\delta u_{0z}}{\gamma_0} + \frac{\left|u_{{\rm f}y}\right|^2}{4\gamma_0^2} \;.
\end{align}
Taking into account that $|u_{{\rm f}y}|=a_0$, which follows from Eq.~\eqref{eq:E}, one finds
\begin{equation}
\label{eq:gammaave}
\frac{\langle \gamma\rangle}{\gamma_0} = 1 - \frac{u_0}{\gamma_0}\frac{\delta u_{0z}}{\gamma_0} + \frac{a_0^2}{4\gamma_0^2} \;.
\end{equation}

The average number density and the average current density can be calculated from Eqs.~\eqref{eq:N}-\eqref{eq:jz}. Taking into account that $n_{\rm f}u_{{\rm f}y}^*+n_{\rm f}^*u_{{\rm f}y} = 0$ and $n_{\rm f}u_{{\rm f}z}^*+n_{\rm f}^*u_{{\rm f}z} = 2n_0u_0 |u_{{\rm f}z}|^2/\gamma_0^2$, which follow from Eqs.~\eqref{eq:nfast}-\eqref{eq:uzfast}, one finds
\begin{align}
\label{eq:Naveaux}
\frac{\langle\gamma n\rangle}{\gamma_0n_0} & = 1 + \frac{\delta n_0}{n_0} - \frac{u_0}{\gamma_0}\frac{\delta u_{0z}}{\gamma_0} + \frac{\left|u_{{\rm f}y}\right|^2}{4\gamma_0^2} - \frac{u_0^2}{\gamma_0^2}\frac{\left|u_{{\rm f}z}\right|^2}{2\gamma_0^2} \\
\label{eq:jyave}
\frac{\langle nu_y\rangle}{\gamma_0n_0} & = \frac{\delta u_{0y}}{\gamma_0}\\
\label{eq:jzaveaux}
\frac{\langle nu_z\rangle}{\gamma_0n_0} & = -\frac{u_0}{\gamma_0} - \frac{u_0}{\gamma_0}\frac{\delta n_0}{n_0} + \frac{\delta u_{0z}}{\gamma_0} + \frac{u_0}{\gamma_0}\frac{\left|u_{{\rm f}z}\right|^2}{2\gamma_0^2} \;.
\end{align}
Taking into account that $|u_{{\rm f}y}|= a_0$ and $|u_{{\rm f}z}|= (\omega_{\rm L}/\omega_{\rm P})|u_{{\rm f}y}|=a_0\omega_{\rm L}/\omega_{\rm P}$, which follow from Eqs.~\eqref{eq:E} and \eqref{eq:uzfast}, one finds
\begin{align}
\label{eq:Nave}
\frac{\langle\gamma n\rangle}{\gamma_0n_0} & = 1 + \frac{\delta n_0}{n_0} - \frac{u_0}{\gamma_0}\frac{\delta u_{0z}}{\gamma_0} + \frac{a_0^2}{4\gamma_0^2} - \frac{u_0^2}{\gamma_0^2}\frac{\omega_{\rm L}^2}{\omega_{\rm P}^2}\frac{a_0^2}{2\gamma_0^2} \\
\label{eq:jzave}
\frac{\langle nu_z\rangle}{\gamma_0n_0} & = -\frac{u_0}{\gamma_0} - \frac{u_0}{\gamma_0}\frac{\delta n_0}{n_0} + \frac{\delta u_{0z}}{\gamma_0} + \frac{u_0}{\gamma_0}\frac{\omega_{\rm L}^2}{\omega_{\rm P}^2}\frac{a_0^2}{2\gamma_0^2} \;.
\end{align}

\end{document}